\documentclass[aps,prb,amsmath,twocolumn]{revtex4}
\usepackage{bm}
\usepackage{graphicx}
\begin{document}
\title{Coulomb Effects on Electron Transport
and Shot Noise in Hybrid Normal-Superconducting Metallic Structures}

\author{Artem V. Galaktionov and Andrei D. Zaikin}
\affiliation{Forschungszentrum Karlsruhe, Institut f\"ur Nanotechnologie,
76021, Karlsruhe, Germany\\
I.E. Tamm Department of Theoretical Physics, P.N.
Lebedev Physics Institute, 119991 Moscow, Russia}

\begin{abstract}
We analyze the effect of electron-electron interactions on Andreev
current and shot noise in diffusive hybrid structures composed of
a normal metal attached to a superconductor via a weakly
transmitting interface. We demonstrate that at
voltages/temperatures below the Thouless energy of a normal metal
Coulomb interaction yields a reduction of both Andreev current and
its noise  by a constant factor which essentially depends
on the system dimensionality. For quasi-1d structures this factor
is $\sim N_{\rm Ch}^{8/g}$, where $N_{\rm Ch}$ and $g$ are
respectively the number of conducting channels and dimensionless
conductance of a normal metal. At voltages above the Thouless
energy the interaction correction to Andreev current and shot
noise acquires an additional voltage dependence which turns out to be a
power-law in the quasi-1d limit.

\end{abstract}
\maketitle

\section{Introduction}
It is well established that low temperature charge transport
through an interface between a normal metal and a superconductor
(NS) is dominated by Andreev reflection \cite{And}: An electron
with energy below the superconducting gap $\Delta$ enters the
superconductor from the normal metal, forms a Cooper pair together
with another electron, while a hole goes back into the normal metal.
In addition to Andreev
reflection electrons can be scattered at the interface potentials
and/or impurities. The combination of ``normal'' and Andreev
reflection mechanisms is essential for proper understanding of
transport phenomena in proximity structures composed of
superconducting and normal metals. This applies both to
equilibrium phenomena, such as dc Josephson effect in diffusive
SNS hybrids \cite{dG,ALO,ZZh,KL,Dubos} and Meissner effect in NS
systems \cite{Z,BBS}, and to non-equilibrium effects, such as
dissipative transport of subgap electrons across NS interfaces \cite{BTK}.

For weakly transmitting NS interfaces the corresponding Andreev conductance
$G_A$ turns out to be rather small as a second order effect in the interface
transmission. At the same time interplay between disorder and interference
effects in the normal metal may strongly enhance Andreev conductance at
sufficiently low energies \cite{VZK,HN,Ben,Zai} leading to the
so-called zero-bias anomaly (ZBA) on the current-voltage characteristics.
In the case of diffusive metals Andreev conductance grows
as
\begin{equation}
G_A \propto 1/\sqrt{T},\;\;\;\; G_A \propto 1/\sqrt{V}
\label{ZBA1}
\end{equation}
with decreasing  temperature and applied voltage $V$.

What is the effect of Coulomb interaction on the Andreev conductance of
NS structures? In the limit of weakly transmitting NS interfaces
this question was addressed in Ref. \onlinecite{Zai} within a
simple capacitive model and the Coulomb blockade of Andreev
reflection was predicted at sufficiently low energies. Huck {\it
et al.} \cite{HHK} studied the problem by modeling the effect of
interactions by means of an effective electromagnetic environment
with an impedance $Z(\omega )$. Similarly to the the case of
normal tunnel junctions \cite{SZ,IN}, for the Ohmic environment
$Z(\omega )=R$ Coulomb interaction yields a power-law ZBA which,
being combined with Eq. (\ref{ZBA1}), yields \cite{HHK}
\begin{equation}
G_A \propto T^{8/g-1/2},\;\;\;\; G_A \propto V^{8/g-1/2}
\label{ZBA2}
\end{equation}
respectively for $eV <T$ and $eV>T$. Here and below we define the
dimensionless conductance as $g=R_q/R$, where $R_q=h/e^2$ is the quantum
resistance unit.

While the approach \cite{HHK} takes into account Coulomb interaction
within a phenomenological scheme of the so-called $P(E)$-theory \cite{SZ,IN},
the question remains if and under which
conditions this scheme is sufficient to account for the effect of
electron-electron interactions in a disordered normal metal attached to a
superconductor. In this paper we develop a microscopic analysis of this
problem and evaluate the interaction correction to Andreev conductance
of diffusive NS structures. We demonstrate that both the form and
the magnitude of this interaction correction essentially depend on the
dimensionality of the sample as well as on the relation between temperature
and applied voltage to the relevant Thouless energy of the structure.
In particular, we predict that at sufficiently low energies the interaction
correction {\it saturates} to a constant value, thus
providing a temperature- and voltage-independent renormalization of
Andreev conductance. This saturation effect was recently
observed \cite{Takayanagi} in NS structures composed of a superconductor and
multi-walled carbon nanotubes (MWNT).

The structure of the paper is as follows. In Sec. II we will
define the model for the NS system and outline the formalism of a
non-linear $\sigma$-model which will be used for our analysis.
Sec. III contains the derivation of the general expression for the
part of the action responsible for Andreev processes in our
system. With the aid of this expression in Sec. IV we will find
the Andreev current in the presence of electron-electron
interactions. Its analysis in a number of important limiting cases
is presented in Sec. V. In Sec. VI we will establish a general
relation between the current and shot noise in the presence of
electron-electron interactions. In Sec. VII we will briefly
discuss our key observations and compare our results with recent
experimental findings \cite{Takayanagi}.

\section{Model and basic formalism}

Throughout the paper we shall consider the hybrid structure depicted in
Fig. 1. A piece of a normal metal of a simple rectangular shape and of
length $L$ is attached to a bulk superconductor via an insulating
interface and to a big metallic reservoir N$'$ via a highly conducting
interface. The transversal dimensions of the normal metal are $L_y$ and
$L_z$, hence, the cross-section of the NS interface $\Gamma$ is just the
product $\Gamma =L_yL_z$. In what follows it will be convenient for us to
assume that the transmission of a tunnel barrier at this interface is
sufficiently low, so that its resistance $R_t$ exceeds that of a normal
metal $R\equiv L/\sigma \Gamma $,
\begin{equation}
R_t \gg R ,
 \label{tN}
\end{equation}
where $\sigma =2e^2DN_0$ is the standard Drude conductivity with
$e$ being the electron charge, $D=v_F l/3$ standing for the diffusion
coefficient, and $N_0$ representing the density of states per spin
direction at the Fermi surface. In addition we
will assume that the normal metal is shorter than the dephasing
length $L_\varphi$,
\begin{equation}
L \ll L_\varphi .
\label{Lphi}
\end{equation}

\begin{figure}
\includegraphics[width=8.cm]{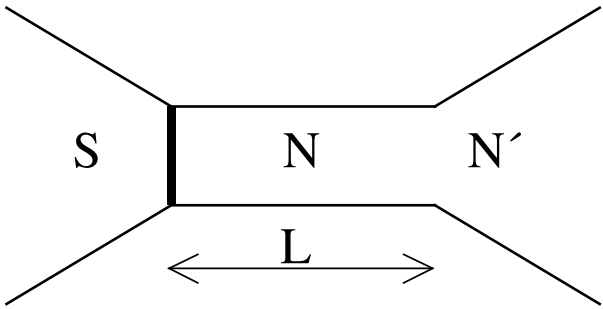}
\caption{NS hybrid structure under consideration.}
\end{figure}

Our theoretical analysis is based  on the Keldysh representation
of a non-linear $\sigma$-model\cite{KA,FLS} which is formulated in
terms of the effective action
\begin{equation}
S=S_0+S_\Gamma ,
\end{equation}
where
\begin{eqnarray}
 && S_0=\frac{i\pi N_0}{4}{\rm Tr}\,\left[ D\left( \bm
\partial \check g \right)^2 + 4i\left(i\check \tau_z
\partial_t+\check\Delta -e\check V\right)\check
g\right], \label{saq2}\\ &&
S_{\Gamma}=-\frac{i\pi}{4e^2R_t\Gamma}{\rm Tr}_\Gamma\,\check
g_-\check g_+. \label{saq3}
\end{eqnarray}
Here $\check g (\bm{R}, t_1,t_2)$ is the $4\times 4$ matrix that
depends on one coordinate and two time variables.  It obeys
the normalization condition
\begin{equation}
\int dt'\check g(\bm{R}, t_1,t') \check g(\bm{R}, t',t_2)=\delta(t_1-t_2).
\label{norm}
\end{equation}
The product of matrices $\check g$ in Eq. (\ref{saq2}) should be
understood as a convolution, cf. Eq. (\ref{norm}). The trace in
Eq. (\ref{saq2}) is taken over the space and time variables and it
is accompanied by the summation over matrix indices.

The term $S_{\Gamma}$ (\ref{saq3}) follows directly from the
Kupriyanov-Lukichev boundary conditions \cite{KL} which account for
tunneling of electrons between N- and S-metals. Spatial integration in
this term is restricted to the insulating interface, subscripts $\mp$
denote matrices taken at the left- and right-hand sides of the interface.
Other matrices in Eq. (\ref{saq2}) are defined as follows
\begin{eqnarray}
&& \check\tau_z=\left( \begin{array}{cc}\hat\tau_z &0\\0& \hat\tau_z
\end{array}\right),\quad \hat\tau_z=\left( \begin{array}{cc}1
&0\\0& -1 \end{array}\right), \label{pmat}
\\ && \check\Delta(\bm{R},t)=\left( \begin{array}{cc}\hat\Delta &0\\0&
\hat\Delta\end{array}\right),\; \hat\Delta= \left( \begin{array}{cc}0
&\Delta(\bm{R},t)\\ -\Delta^*(\bm{R},t)&0
\end{array}\right).
\nonumber
\end{eqnarray}
The operator $\bm\partial $ acts as
\begin{eqnarray}
&& \bm\partial \check g= \nabla\check g-i\frac{e}{c}
\left[\bm{A}\check\tau_z,\check g\right]\equiv \nonumber\\ &&
\nabla\check g(t,t') -i\frac{e}{c} \bm{A}(t)\check\tau_z \check
g(t,t')+\check g(t,t') i\frac{e}{c} \bm{A}(t')\check\tau_z.
\label{partial}
\end{eqnarray}
Here $\bm{A}(\bm{R},t)$ is the vector potential. In Eq.
(\ref{partial}) spatial arguments are suppressed for brevity.

Finally, the $\check V$ term depends on the Hubbard-Stratonovich
fields $V_1$ and $V_2$ defined on the two branches of the Keldysh
contour.  These fluctuating fields emerge after the standard
decoupling procedure in the term describing electron-electron
interactions. We define
\begin{equation}
\check V(\bm{R},t)=\left(\begin{array}{cc}V^+\hat 1& \frac{1}{2}
V^-\hat 1
\\ \frac{1}{2} V^-\hat 1 & V^+\hat 1
\end{array} \right),
\end{equation}
where $\hat 1$ is $2\times2$ unity matrix and $V^+=(V_1+V_2)/2,\,
V^-=V_1-V_2$ define respectively "classical" and "quantum"
fluctuating fields. If an external voltage $V$ is applied to the
system, it should simply be added to the classical component
$V^+$.

It is necessary to supplement the action (\ref{saq2}),
(\ref{saq3}) describing fermionic degrees of freedom, by the
action for the bosonic fields $V^\pm$. The latter action
determines the correlators for these fields which will be
specified below.

\section{Andreev action}

Let us first define the matrices $\check g_{\pm}$ in the absence of both
electron-electron interactions and tunneling between metals. For the
normal metal one has $\check g_+(\bm{R},t,t')=\check g_0(t,t')$, where
\begin{eqnarray} && \check g_0=\left(\begin{array}{cc} \hat g^R_0&
\hat g^K_0\\ 0& \hat g^A_0
\end{array}\right),\quad \hat
g^{R,A}_0(t,t')=\pm\hat\tau_z\delta(t-t'), \nonumber\\ && \hat
g^{K}_0(t,t')=-\frac{2i T}{\sinh[ \pi T(t-t')]}
\hat\tau_z.\label{equil}
\end{eqnarray}
The Fourier transform of the Keldysh matrix $\hat g^{K}_0$ reads
$\hat g^{K}_0(\epsilon)=2\tanh[\epsilon/2T] \hat\tau_z$. In what
follows we will consider the values of voltages and temperatures
much smaller than the superconducting gap $eV,T \ll \Delta$.
Hence, it will be sufficient for our purposes to specify the
matrix $\check g_-$ for a superconducting electrode only at
energies much smaller than $\Delta$. Under this condition we have
\begin{equation}
\check g_-(t,t')=\left(\begin{array}{cccc}0& -i&0 &0\\ i&0 &0 &0\\0&
0&0&-i\\ 0& 0 &i &0
\end{array} \right)\delta(t-t').
\end{equation}

Let us turn on electron-electron interactions. While in a
large superconducting electrode their influence on the matrix
$\check g_-$ can be safely neglected, the matrix $\check g_+$ in
the normal metal gets modified as \cite{SZ,KA}
\begin{equation}
\check g_+(\bm{R},t_1,t_2)=e^{-i\check K(\bm{R},t_1)} \check
g_0(t_1,t_2)e^{i\check K(\bm{R}, t_2)}. \label{gt}
\end{equation}
Here the matrix $\check K$ has the structure
\begin{equation}
\check K=\left(\begin{array}{cc} \varphi^+ \hat\tau_z&
\varphi^-\hat\tau_z/2\\ \varphi^-\hat\tau_z/2&  \varphi^+ \hat\tau_z
\end{array}\right), \label{sf}
\end{equation}
and the phase fields $\varphi^{\pm}$ satisfy the equations
\begin{eqnarray}
&& (Dq^2 + i\omega)\varphi^-(\bm{q},\omega)+eV^-(\bm{q}, \omega)=0,
\nonumber
\\&& (Dq^2 -i\omega) \varphi^+(\bm{q},\omega)-eV^+(\bm{q},
\omega)=\nonumber\\&& -Dq^2\varphi^-(\bm{q},\omega)\coth\frac{\omega}{2T}.
\label{phdef}
\end{eqnarray}
One can easily check that the part of the variation of $S_0$
(\ref{saq2}) linear both in $\check g$-variations and in the
fields $V^\pm$ vanishes under the transformation
(\ref{gt})-(\ref{phdef}). These equations are sufficient to
account for the effect of Coulomb interactions for the problem
in question.

Now let us include tunneling between metals into consideration. In
order to eliminate the dependence of the matrix $\check g_+$ on
the fluctuating fields let us permute the factors $\exp(\pm
i\check K)$ in the term $S_\Gamma $ (\ref{saq3}) to act on the
matrix $\check g_-$. After this transformation we obtain
\begin{equation}
S_\Gamma=-\frac{i\pi}{4e^2R_t \Gamma} {\rm Tr}_\Gamma\,\check Q_-\check
g_+, \label{bt}
\end{equation}
where
\begin{eqnarray}
&& \check Q_-(\bm{R},t, t')=\left(\begin{array}{cc} \hat A(\bm{R},t)& \hat
B(\bm{R},t) \\  \hat B(\bm{R},t)& \hat A(\bm{R},t)
\end{array}\right)\delta(t-t'),\nonumber\\
&& \hat A=\left(\begin{array}{cc} 0& -ie^{2i\varphi^+}\cos \varphi^-
\\ie^{-2i\varphi^+}\cos\varphi^-& 0\end{array}\right),\label{mqs}\\
&& \hat B=\left(\begin{array}{cc} 0& e^{2i\varphi^+}\sin \varphi^-
\\e^{-2i\varphi^+}\sin\varphi^-& 0\end{array}\right).\nonumber
\end{eqnarray}

What remains is to evaluate the deviations of $\check g_+$ from
its normal state value (\ref{equil}) due to tunneling of Cooper
pairs from the superconducting electrode into the normal metal. It
is convenient to employ the following parameterization
\begin{equation} {\cal Q}=\check u\check g_+\check u, \quad
\check u=\check u^{-1}= \left(\begin{array}{cc} \delta(t-t')\hat 1&
-\frac{i T}{\sinh[ \pi T(t-t')]}\hat 1 \\ 0 & -\delta(t-t')\hat 1
\end{array}\right),\label{mdiag}
\end{equation}
which brings the matrix $\cal Q$ to the diagonal form in the absence of the
proximity effect. The latter effect can can then be interpreted in terms of
fluctuations of  $\cal Q$ parameterized as
\begin{equation}
{\cal Q}=e^{-{\cal W}/2}\check\sigma_z\check\tau_z  e^{{\cal
W}/2}=\check\sigma_z\check\tau_z \left(1+{\cal W}+{\cal
W}^2/2+\ldots\right)
\label{WWW}
\end{equation}
with
\begin{equation}
\check\sigma_z=\left( \begin{array}{cc}\hat 1 &0\\0&-\hat 1
\end{array}\right).
\end{equation}
The matrix ${\cal W}$ is in turn parameterized via the Pauli
matrices $\hat\tau_{x,y,z}$ as
\begin{equation}
{\cal W}=\left( \begin{array}{cc} w_1\hat\tau_x+w_2\hat\tau_y & w_0\hat 1+
w_3\hat\tau_z
\\ \overline w_0\hat 1+ \overline w_3\hat\tau_z &
\overline w_1\hat\tau_x+\overline w_2\hat\tau_y\end{array} \right).
\end{equation}
The functions $w_i$, and $\overline w_i$ can be expanded in
the eigenfunctions $\psi$ of the diffusion equation
\begin{equation}
w_i(x,y,z)=\sum_{\bm{q}} w_i(\bm{q})\psi_{\bm{q}}(x,y,z),
\end{equation}
obeying the boundary conditions $d\psi /dx|_{x=0}=0$, $\psi
|_{x=L}=0$ and impenetrable boundary conditions at $y= 0, L_y$ and
$z= 0, L_z$. We choose
\begin{eqnarray}
&& \psi_{\bm{q}}(x,y,z)\propto \cos q_x x  \cos q_y y \cos q_z z,
\nonumber\\ && q_x=\frac{\pi}{L}\left( l+\frac{1}{2}\right),\;
q_y=\frac{\pi}{L_y}m ,\; q_z=\frac{\pi}{L_z}n,\nonumber\\ && l,m,n=0,1,...
\label{dc2}
\end{eqnarray}
The normalization constant is determined by the condition $\int dx dy dz
\psi_{\bm{q}}^2(\bm{r})=1$.

Using the effective action (\ref{saq2}), we can find all non-zero
averages for the products of the functions $w_i$, and $\overline
w_i$. Below we will only need the following averages \cite{FLS}
\begin{eqnarray}
&& \left\langle w_i(\bm{q},\epsilon_1,\epsilon_2)w_i(-
\bm{q},\epsilon_3,\epsilon_4) \right\rangle=\nonumber\\&& -\frac{1}{\pi
N_0}\frac{(2\pi)^2\delta(\epsilon_1-\epsilon_4)\delta(\epsilon_2-
\epsilon_3)}{Dq^2-i(\epsilon_1+\epsilon_2)},\nonumber\\ && \left\langle
\overline w_i(\bm{q},\epsilon_1,\epsilon_2)\overline w_i(-
\bm{q},\epsilon_3,\epsilon_4) \right\rangle=\nonumber\\&& -\frac{1}{\pi
N_0}\frac{(2\pi)^2\delta(\epsilon_1-\epsilon_4)\delta(\epsilon_2-
\epsilon_3)}{Dq^2+i(\epsilon_1+\epsilon_2)},\label{avp}
\end{eqnarray}
which correspond to the Cooperons $(i=1,2)$.

Now we are ready to evaluate the contribution of Andreev processes
to the effective action. In the weak tunneling limit it is
sufficient to expand $\exp (iS )$ up to the second order in
$S_\Gamma$ and then to average the resulting expression over the
fermionic degrees of freedom with subsequent re-exponentiation of
the result. In the limit $R_t\gg R$, the amplitude of the
anomalous Green function penetrating into the normal metal remains
much smaller than unity. Hence, in Eq. (\ref{WWW}) it is
sufficient to retain only the linear terms in ${\cal W}$. The
resulting Andreev contribution to the action then takes the form
\begin{eqnarray}
&& \delta S_A=-\frac{i}{32}\left( \frac{\pi}{e^2 R_t
\Gamma}\right)^2\label{sot}\\ && \left\langle{\rm Tr}_{\Gamma_1}
{\cal Q}_- ( \Gamma_1)\check \sigma_z\check\tau_z{\cal
W}(\Gamma_1)\cdot {\rm Tr}_{\Gamma_2} {\cal Q}_- (
\Gamma_2)\check\sigma_z\check\tau_z{\cal
W}(\Gamma_2)\right\rangle.\nonumber
\end{eqnarray}
The space integration is performed here along the same interface
$\Gamma$, but independently for the terms containing $\Gamma_1$ and $\Gamma_2$.
The matrix ${\cal Q}_-$ is defined by the relation
${\cal Q}_-=\check u \check Q_- \check u$. Employing Eqs. (\ref{mqs}),
(\ref{mdiag}) and performing the averaging with the aid of Eqs. (\ref{avp}),
we arrive at the final expression for $\delta S_A [\varphi^\pm ]$ which will
be extensively used in our subsequent analysis. We find

\begin{widetext}
\begin{eqnarray}
&& \delta S_A=\frac{i\pi R D}{2e^2R_t^2\Gamma L}\sum_{\bm{q}}\int\frac{d E
d\omega }{(2\pi)^2}\int_\Gamma d\bm{r}\int_\Gamma d\bm{r'}\int d t\int d
t'
e^{i\omega(t-t')}\psi_{\bm{q}}(\bm{r})\psi_{\bm{q}}(\bm{r'})\times\nonumber\\
&& \bigg\{
\frac{Dq^2}{(Dq^2)^2+4E^2}\left[\tanh\frac{E+(\omega/2)}{2T}-
\tanh\frac{E-(\omega/2)}{2T}
\right]
\bigg[\coth\frac{\omega}{2T}e^{-2i(\varphi^+(\bm{r},t)-\varphi^+(\bm{r'},t'))}
\sin[\varphi^-(\bm{r},t)] \sin[\varphi^-(\bm{r'},t')] \nonumber\\ &&+
\sin\left[2\varphi^+(\bm{r},t)-2\varphi^+(\bm{r'},t')\right]
\cos[\varphi^-(\bm{r},t)] \sin[\varphi^-(\bm{r'},t')] \bigg]\nonumber
\\
&& -\frac{2iE}{(Dq^2)^2+4E^2}\left[\tanh\frac{E+(\omega/2)}{2T}+
\tanh\frac{E-(\omega/2)}{2T} \right]
\sin\left[2\varphi^+(\bm{r},t)-2\varphi^+(\bm{r'},t')\right]
\cos[\varphi^-(\bm{r},t)]\sin[\varphi^-(\bm{r'},t')]\nonumber
\\&& -\frac{Dq^2}{(Dq^2)^2+4E^2}e^{2i(\varphi^+(\bm{r},t)-
\varphi^+(\bm{r'},t'))}
\cos[\varphi^-(\bm{r},t)-\varphi^-(\bm{r'},t')]\bigg\}. \label{finalaction}
\end{eqnarray}
\end{widetext}

\section{Andreev current in the presence of interactions}

Let is first evaluate Andreev current across the NS interface.
Substituting Eq. (\ref{finalaction}) into the formal expression for the current
\begin{equation}
I=e \left\langle \int_\Gamma d^2 r\frac{\delta}{\delta
\varphi^-(\bm{r})} \delta S_A \right\rangle
\label{curdef}
\end{equation}
and
performing averaging over the fluctuating fields $\varphi^\pm$ we
arrive at the general result for the Andreev current
\begin{widetext}
\begin{eqnarray}
&&  I=\frac{\pi R D}{2e R_t^2\Gamma L}\sum_{\bm{q}}\int\frac{dE
d\omega}{(2\pi)^2}\int_{\Gamma} d^2 r d^2 r'\psi_{\bm{q}}(\bm{r})
\psi_{\bm{q}}(\bm{r'}) \left[ \tanh\left(
\frac{E+eV-(\omega/2)}{2T}\right)-  \tanh\left(
\frac{E-eV+(\omega/2)}{2T}\right)\right]\times \nonumber
\\&&
\frac{Dq^2}{\left(Dq^2\right)^2+4E^2} P(\omega,\bm{r},
\bm{r'})\frac{1-e^{-2eV/T}}{1-e^{(\omega-2eV)/T}}. \label{main}
\end{eqnarray}
\end{widetext}
Here the function
\begin{eqnarray}
&& P(\omega,\bm{r},\bm{r'})=\int dt e^{J(t,\bm{r},\bm{r'}) +i\omega
t},\nonumber
\\ && e^{J(t,\bm{r},\bm{r'})}=  \left\langle e^{2i\varphi_{2}(\bm{r},t)-
2i\varphi_{1}(\bm{r'},0)} \right\rangle \label{J}
\end{eqnarray}
accounts for the effect of interactions described by the fluctuating phase
fields $\varphi_{1,2} =\varphi^+\pm (\varphi^-/2)$ defined on the two
branches of the Keldysh contour. Averaging in Eq. (\ref{J}) is
performed in the absence of the external voltage. Note that the form of
this function is reminiscent to that used in the standard $P(E)$-theory
\cite{SZ,IN} and the result for the Andreev current (\ref{main}) looks
somewhat similar to the expression derived within the framework of the
latter theory in Ref. \onlinecite{HHK}. In contrast to the latter work,
however, our approach and resulting Eqs. (\ref{main}), (\ref{J}) fully
account (i) for electron
coherence across the whole NS structure and (ii) for the spatial
dependence, introduced both by the functions $\psi_{\bm{q}}(\bm{r})$,
describing fermionic fluctuations, and by the function
$J(t,\bm{r},\bm{r'})$, describing bosonic (phase) fluctuations.

These fluctuations depend on the dimensionality of the normal
sample. For our purposes it will be sufficient to employ the
standard random phase approximation (RPA)
and to express the correlators  of the fluctuating
Hubbard-Stratonovich fields via an effective $d$-dimensional
dielectric function $\varepsilon(\omega,k)$ of our system. One
finds \cite{GZ}
\begin{eqnarray}
&& \left\langle e V^+(t,\bm{r}) e
V^+(0,0)\right\rangle=\nonumber\\ && -\int \frac{d\omega d^d
k}{(2\pi)^{d+1}}{\rm Im}\left( V^*(k,\omega) \right)\coth
\left(\frac{\omega}{2T}\right) e^{-i\omega t+i\bm{k
r}},\nonumber\\ && \left\langle e V^+(t,\bm{r}) e
V^-(0,0)\right\rangle=\nonumber\\ && i\int \frac{d\omega d^d
k}{(2\pi)^{d+1}}V^*(k,\omega) e^{-i\omega t+i\bm{k r}}\nonumber,\\
&& \left\langle e V^-(t,\bm{r}) e
V^-(0,0)\right\rangle=0.\label{aver}
\end{eqnarray}
Here we have defined
\begin{equation}
V^*(k,\omega)=\frac{1}{V_0^{-1}(k)+\frac{k^2}{4\pi
e^2}\left[\varepsilon(k,\omega)-1\right]},\label{vstar}
\end{equation}
where the term $V_0(k)$ stands for the unscreened Coulomb
interaction:
\begin{eqnarray}
V_0(k)=\frac{4\pi e^2}{k^2},\quad d=3,\nonumber
\\ V_0(k)=\frac{2\pi e^2}{k},\quad d=2,\nonumber
\\V_0(k)=e^2\ln\left(1+\frac{1}{\Gamma k^2}\right),\quad d=1.
\label{usc}
\end{eqnarray}
In the latter equation we have assumed $L_y\sim L_z \sim
\sqrt{\Gamma}$. In the three-dimensional case we obviously have
$V^*(k,\omega)=V_0(k)/\varepsilon(k,\omega)$.

Note, that in the case of lower dimensions $d=1,2$ the term
$V_0^{-1}(k)$ also accounts for the electric field energy outside
the conductors. The corresponding interaction can be screened,
provided there are metallic gates near the film/wire. For
instance, in the case $d=2$ one has
\begin{equation}
V_0^{scr}(k)=\frac{2\pi e^2}{k}\left(1-e^{-bk}\right),
\end{equation}
where $b$ is the distance between the film and the gate electrode.

In order to evaluate the average for the phase fields (\ref{J})
one can combine Eqs. (\ref{phdef}), (\ref{aver}) with the standard
Drude form of the dielectric function for the normal metal
\begin{equation}
\varepsilon(k,\omega)=1+\frac{4\pi \sigma}{-i\omega+Dk^2}. \label{rpa}
\end{equation}

The resulting expressions for the function (\ref{J}) obtained in this way
remain applicable at frequencies above the corresponding Thouless energy
of the sample but should fail at lower frequencies, in which case
discrete wave vectors $\sim 1/L$ become significant. In order to
cure this complication it is
useful to depart from the representation of plane waves in Eq.
(\ref{aver}) and to employ a different set of the (volume-normalized) basis
functions $\Omega_{m}(\bm{r})$ obeying the equation
\begin{equation}
\nabla^2 \Omega_{m}(\bm{r})+ \lambda_m^2\Omega_{m}(\bm{r})=0
\end{equation}
with the corresponding eigenvalues $\lambda_m$. At the
impenetrable interfaces the functions $\Omega_{m}(\bm{r})$ satisfy
the boundary condition $\nabla_n \Omega_{m}(\bm{r})=0$. One should
also require the functions
$\Omega_{m}(\bm{r})$ to vanish at the interface with a large
metallic reservoir $x=L$.

The corresponding generalization of the dielectric function
(\ref{rpa}) can be achieved by evaluating the response of a finite sample,
characterized by  the current density $\bm{j}(\bm{r})=\sigma
\bm{E}(\bm{r})-D\nabla \rho(\bm{r})$ with $\bm{E}$ and $\rho$ standing
respectively for the
electric field and for the charge density. One gets \cite{SG}
\begin{eqnarray}
&& \varepsilon_{\alpha
\beta}(\bm{r},\bm{r'},\omega)=\delta_{\alpha\beta}\delta(\bm{r}-\bm{r'})
+\\ && \frac{4\pi i}{\omega}\sigma_D
\left(\delta_{\alpha\beta}\delta(\bm{r}-\bm{r'})-\nabla_{\alpha}\nabla'_{\beta}
d(\bm{r},\bm{r'},\omega)\right), \nonumber
\end{eqnarray}
where
\begin{equation}
d(\bm{r},\bm{r'},\omega)=\sum_{m}\frac{\Omega_{m}(\bm{r})
\Omega_{m}(\bm{r'})}{\lambda_m^2-\frac{i\omega}{D}}.
\end{equation}

Implementing these modifications also into Eqs. (\ref{aver}) and employing
Eqs. (\ref{phdef}) we arrive at the following general result
\begin{eqnarray}
&& J(t,\bm{r},\bm{r'})=4\sum_{m}\int\frac{d\omega }{(2\pi)}\,{\rm
Im}\left[ \frac{V^*(\lambda_m,\omega)}{
(D\lambda_m^2-i\omega)^2}\right]\times\nonumber\\&&
\bigg[\left(\coth\frac{\omega}{2T} \cos \omega t -i \sin \omega
t\right)\Omega_m(\bm{r})\Omega_m(\bm{r'})- \label{jdf2}\\ &&
\frac{1}{2}\coth\frac{\omega}{2T} \left( \Omega_m(\bm{r})\Omega_m(\bm{r})+
\Omega_m(\bm{r'})\Omega_m(\bm{r'}) \right)\bigg]. \nonumber
\end{eqnarray}
The latter formula results in
$J(t-(i/T),\bm{r},\bm{r'})=J(-t,\bm{r},\bm{r'})$ leading to the relation
$P(-E,\bm{r},\bm{r'})=e^{-E/T}P(\omega,\bm{r},\bm{r'})$ as in the standard
$P(E)$ theory. Accordingly, at $T \to 0$ Eq. (\ref{main}) reduces to
\begin{eqnarray}
&& \frac{dI}{dV}=\frac{D R}{2\pi R_t^2\Gamma L}
\sum_{\bm{q}}\int\limits_{0^-}^{2|eV|}dE \int_{\Gamma} d^2 r d^2
r'\psi_{\bm{q}}(\bm{r}) \psi_{\bm{q}}(\bm{r'})\nonumber \\ && P(E,\bm{r},
\bm{r'}) \frac{Dq^2}{\left(Dq^2\right)^2+(2|eV|-E)^2}. \label{ea}
\end{eqnarray}
 At frequencies exceeding the sample Thouless energy it is possible to
perform averaging of Eq. (\ref{jdf2}) over fluctuations depending on
$\bm{r}+\bm{r'}$. Then one gets
\begin{eqnarray}
&& J(t,|\bm{r}|)=4\int\frac{d\omega d^n k}{(2\pi)^{n+1}}\,{\rm Im}\left[
\frac{V^*(k,\omega))}{ (Dk^2-i\omega)^2}\right]\times\nonumber\\&&
\left[\coth\frac{\omega}{2T} (\cos \omega t\cos\bm{k}\bm{r}-1 )-i \sin
\omega t\cos\bm{k}\bm{r}\right]. \label{jdf}
\end{eqnarray}

\section{Results}

\subsection{Non-interacting limit}

Let us first reproduce the results for Andreev current in the
non-interacting limit. In this case we have $P(\omega,r)=2\pi
\delta(\omega)$ and the standard expressions for the $I-V$ curve
are readily recovered. In the limit $eV \gg T$ one finds
\begin{eqnarray}
&& \frac{dI}{dV}=\frac{R}{R_t^2}, \quad {\rm if}\;
|eV|\ll \epsilon_{\rm Th}; \nonumber
\\ && \frac{dI}{dV}=\frac{R}{2R_t^2}\sqrt{\frac{\epsilon_{\rm Th}}{|eV|}}
, \quad {\rm if}\; |eV| \gg
\epsilon_{\rm Th}.
\label{ZBA}
\end{eqnarray}
Here and below $\epsilon_{\rm Th}=D/L^2$ is the Thouless energy.

In the linear regime $eV \ll T$ for the Andreev conductance $G_A\equiv dI/dV$
we obtain
\begin{eqnarray}
&& G_A^{(0)}=\frac{R}{R_t^2}, \quad {\rm if}\;
T\ll \epsilon_{\rm Th},
\label{ZBA0}
\\ && G_A^{(0)}=\frac{(2^{3/2}-1)\zeta(3/2)}{4\sqrt{\pi}}\frac{R}{R_t^2}
\sqrt{\frac{\epsilon_{\rm Th}}{T}}
, \quad {\rm if}\; T  \gg \epsilon_{\rm Th},\nonumber
\end{eqnarray}
where $(2^{3/2}-1)\zeta(3/2)/4\sqrt{\pi}\approx 0.67$.

Now let us turn to the effect of electron-electron interactions.
Both the value and the form of the interaction correction to the
Andreev conductance essentially depend on the effective
dimensionality of fluctuations of the phase fields. Depending on
the voltage and temperature the leading contribution to the
interaction correction is given either by the modes with $l, m,
n\gg 1$ (cf. Eq. (\ref{dc2})) or by the modes with $m$ (or $n$, or
both) being equal to zero. Correspondingly, below we will
distinguish between effectively 3d-, quasi-2d- and quasi-1d-type
of behavior of the interaction correction.

\subsection{Effect of interactions: Bulk limit}

We start from the case of 3d fluctuations. For convenience we shall assume
$L\gtrsim L_y \gtrsim L_z$ and consider the limit $eV \gg T$.
At sufficiently
high voltages $eV\gg D/L_z^2$ from Eq. (\ref{jdf}) (with $d=3$) we obtain
\begin{eqnarray}
&& J(t,r)=\frac{e^2}{\pi^2 \sigma r}\int\limits_0^\infty \frac{\cos
(2Dt\omega/r^2)}{\omega}\left(
1-e^{-\sqrt{\omega}}\cos\sqrt{\omega}\right) d\omega\nonumber\\ &&
-\frac{ie^2}{2\pi\sigma r} {\rm erf}\left(\frac{r}{2\sqrt{D|t|}}
\right){\rm sgn}\, t -\kappa ,
\end{eqnarray}
where $ {\rm erf}(x)=2\int_0^x \exp(-t^2) d t/\sqrt{\pi}$ and
\begin{equation}
\kappa=4\int\frac{d\omega d^3 k}{(2\pi)^{4}}\,{\rm Im}\left[
\frac{V^*(k,\omega)}{ (Dk^2-i\omega)^2}\right]\rm{sign}\,\omega.
\end{equation}
Making use of the condition $P(\omega, r)=0$ for $\omega <0$, we have for
$\omega\ge 0$
\begin{eqnarray}
&& P(\omega, r)=e^{-\kappa} 2\pi\delta(\omega) +\nonumber\\ &&\frac{2e^2
e^{-\kappa} }{\pi\sigma r\omega}\left[ 1-e^{-r\sqrt{\omega/2D}}\cos \left(
r\sqrt{\omega/2D}\right)\right]. \label{per}
\end{eqnarray}
Thus we obtain
\begin{equation}
\frac{dI}{dV}=\frac{Re^{-\kappa }}{2 R_t^2}\left(\sqrt{\frac{\epsilon_{\rm
Th}}{|eV|}} +\frac{3.64 \Gamma}{g L^2} \right). \label{as3}
\end{equation}

One observes that the relative ``weight'' of the second term in the
brackets of (\ref{as3}) grows with voltage as
$(1/g)\sqrt{eV/\epsilon_{\rm Th}}$. This form is consistent with
the standard square-root energy dependence of the interaction
correction to the density of states. In the second term in Eq.
(\ref{as3}) this dependence is exactly compensated by ZBA
(\ref{ZBA}) resulting in the voltage-independent contribution
$\sim R/gR_t^2$. This compensation will not occur in other limits
to be considered below.

The value of $\kappa$ in the bulk limit can be estimated as
\begin{equation}
\kappa \equiv \kappa_3 \sim \frac{e^2}{\sigma
\sqrt{D}}\int\limits_0^{1/\tau}\frac{d\omega}{\sqrt{\omega}}\sim
\frac{1}{(p_Fl)^2}.\label{33}
\end{equation}
Here $\tau=l/v_F$ is the elastic scattering time.

While the result (\ref{as3}) holds at high voltages, at lower
values $eV \lesssim \epsilon_{\rm Th}$ -- as it was already
discussed above -- the Andreev conductance saturates to a constant
value
\begin{equation}
\frac{dI}{dV}=\frac{Re^{-\kappa}}{R_t^2}\label{satt}
\end{equation}
with the voltage-dependent correction to this formula of order
$(eV)^2/g\epsilon_{\rm Th}^2$.

Note that in the bulk limit the net effect of
electron-electron interactions is parametrically small and negative in the
admissible range of voltages $eV < 1/\tau$. Indeed, having in mind that
our calculation is performed in the limit $p_Fl \gg 1$ and making use of
the estimate (\ref{33}), in Eq. (\ref{as3}) one can expand $\exp (-\kappa
)$ to the first order in $\kappa$ and observe that the corresponding
negative correction exceeds the positive one by a factor $\sim
1/\sqrt{|eV|\tau}$.

Let us also point out that in order to obtain the Andreev
conductance in the voltage range $eV \ll T$ it is sufficient to
simply substitute $T$ instead of $eV$. As in the non-interacting
limit (cf. Eqs. (\ref{ZBA}) and (\ref{ZBA0})), this procedure will
yield the correct form of the interaction correction up to an
unimportant numerical prefactor of order one.

\subsection{Effect of interactions: Quasi-2d structures}

Let us now turn to low-dimensional fluctuations. We first consider
systems which can be treated as quasi-2d ones, i.e. we assume that
both the length and the width of the normal metal strongly exceed
its thickness $L\sim L_y \gg L_z$. We shall address two different
physical situations corresponding to the absence and to the
presence of massive metallic gate electrodes in the vicinity of
the sample.

In the absence of the gate electrodes we can use the 2d expression
for $V_0(k)$ (\ref{usc}), combined with Eq. (\ref{vstar}). We
first evaluate the constant $\kappa$ which now consists of two
contributions,
\begin{equation}
\kappa = \kappa_3 +\kappa_2,
\end{equation}
where the first one is again determined by high frequencies ($\omega
\gtrsim D/L_z^2$) and small scales (cf. Eq. (\ref{33})), while the second
one emerges from integrating over the parameter region $ 2\pi\sigma L_z
q\gg \omega$ and $Dq^2\ll \omega$ in Eq. (\ref{jdf2}) for smaller
frequencies $\omega \lesssim D/L_z^2$
\begin{equation}
\kappa_2 =\frac{8 L_y}{\pi g L}\ln\frac{L_y}{L_z}\ln\frac{\omega_0
\Gamma}{D}, \quad \omega_0=\frac{(2\pi \sigma L_z)^2}{D}.
\end{equation}
It is obvious from the above expressions that the contribution
$\kappa_3$ exceeds $\kappa_2$ for thicker films, i.e. as long as
$L_z$ remains bigger than the mean free path $l$. In the opposite
case, however, the contribution of 3d fluctuations $\kappa_3$ is
negligible, and in the expression for $\kappa_2$ one should
substitute $1/\tau$ instead of $D/L_z^2$ in the arguments of the
logarithms. Then one finds
\begin{equation}
\kappa \simeq \kappa_2 =\frac{2 L_y}{\pi g L}
\ln\frac{L_y^2}{D\tau}\ln\frac{\omega_0^2 \tau
L_y^2}{D}.
\label{22}
\end{equation}
We observe that the above expressions for the renormalization
constant $\kappa$ contain double logarithmic factors which are
rather typical for 2d structures. Similar double logarithmic
dependencies emerge in voltage- and temperature-dependent
contributions to be analyzed below.

Let us again stick to the limit $eV \gg T$. At voltages
$D/L_y^2\ll eV\ll D/L_z^2$ we obtain
\begin{equation}
\frac{dI}{dV}=\frac{R}{2R_t^2}\sqrt{\frac{\epsilon_{\rm
Th}}{|eV|}}e^{-\kappa}
 \left[1+ \frac{2 L_y}{\pi g L}\ln
\frac{|eV| L_y^2}{D} \ln\frac{\omega_0^2L_y^2}{|eV|D}\right].\label{2dng}
\end{equation}
Note that in contrast to the expression for $\kappa$ (\ref{22})
the voltage-dependent part of the interaction correction
(\ref{2dng}) does not involve the upper frequency cutoff $\sim
1/\tau$. At lower voltages $eV\lesssim \epsilon_{\rm Th}$ the
differential conductance again saturates to the value (\ref{satt})
with $\kappa$ defined in Eq. (\ref{22}). The above results do not
coincide with ones obtained in Ref. \onlinecite{FLS} for the same
physical situation.

It is worth pointing out that even though the combination $L_y/gL$ in the
above expressions always remains small, in some cases this smallness can be
compensated by large logarithms, and the formal expansion in $1/g$ can become
insufficient. In this situation the $I-V$ curve can be evaluated with the aid
of the following expression
\begin{equation}
P(E)=2\pi\delta(E)e^{-\Phi(0)} +2\pi \frac{d}{dE}e^{-\Phi(E)},
\end{equation}
where
\begin{equation}
\Phi(E)=\frac{e^2}{\pi^2\sigma L_z
}\ln\frac{1}{\sqrt{\epsilon_{\rm Th}^2+E^2}\tau}
\ln\frac{\omega_0^2 \tau}{\sqrt{\epsilon_{\rm Th}^2+E^2}}.
\end{equation}
This function also enters into the expression for the density of states
of normal 2d films \cite{KA}
\begin{equation}
N(\epsilon)=N_0 e^{-\Phi(E)/4}.
\end{equation}
We also note that in the case of thicker films $L_z \gg l$ our results
are consistent with the expression for the density of states \cite{Grab}.

In the presence of massive gate electrodes close to our sample
the fluctuating electromagnetic fields do not extend outside its volume, and
the term $V^*$ in Eq. (\ref{jdf2}) gets modified. In this limit we find
\begin{equation}
\frac{dI}{dV}=\frac{R}{2R_t^2}\sqrt{\frac{\epsilon_{\rm
Th}}{|eV|}}e^{-\kappa}
 \left[1+ \frac{ 2 L_y}{\pi g L}\ln^2
\frac{|eV| L_y^2}{D}\right], \label{f2}
\end{equation}
where
\begin{equation}
\kappa= \frac{ 2 L_y}{\pi g L}\ln^2 \frac{L_y^2}{\tau D}
\end{equation}

If one considers the limit $T \gg eV$, it suffices to substitute $0.56 T$
instead of $eV$ in Eqs. (\ref{2dng}), (\ref{f2}) , and one will arrive at
the expression for the linear Andreev conductance $G_A(T)$ valid with the
logarithmic accuracy.

\subsection{Effect of interactions: Quasi-1d structures}

Let us finally turn to quasi-1d structures $L \gg L_y\sim L_z$ and
examine the effect of electron-electron interactions on the
Andreev conductance in the absence of screening gates. Similarly
to the 2d case, there also exists the parameter region $Dq^2\ll
\omega$ and $\sigma \Gamma q^2\log(1/L_z^2q^2)\gg \omega$ , which
gives the leading contribution to $J(t)$. We obtain
\begin{eqnarray}
&& J(t)=16\int\limits_{\epsilon_{\rm Th}}^\infty
\int\limits_{q_{\rm min}}^\infty \frac{dq}{2\pi}\frac{e^2}{\sigma
\Gamma \omega q^2}\times\nonumber\\ &&
\left\{\coth\frac{\omega}{2T}\left[ \cos(\omega t)-1 \right]
-i\sin(\omega t)\right\}.
\end{eqnarray}
Since $1/q$ cannot exceed $L$, one should choose
\begin{equation}
q_{\rm min}\sim L^{-1}+\sqrt{\frac{\omega}{\sigma \Gamma\ln[\sigma
\Gamma/\omega L_z^2]}}.
\end{equation}
Performing the $q$-integration one finds
\begin{eqnarray}
&& J(t)=\frac{8}{g}\int\limits_{\epsilon_{\rm
Th}}^\infty\frac{d\omega}{\omega\left(1+ \sqrt{\omega
RC}\right)}\nonumber \\ && \left\{\coth\frac{\omega}{2T}\left[
\cos(\omega t)-1 \right] -i\sin(\omega t)\right\}, \label{mpe}
\end{eqnarray}
where
\begin{equation}
C=\frac{L}{\ln[L^2/\Gamma ]}
\label{Cap}
\end{equation}
is the capacitance of a long cylinder. The expression (\ref{mpe})
is rather similar to that obtained within the
$P(E)$-theory\cite{HHK}. There are at least two differences,
however. One of them is that here the combination $1+\sqrt{\omega
RC}$ enters in the denominator of Eq. (\ref{mpe}), while  in the
model \cite{HHK} one has $1+\omega^2R^2C_J^2$, where $C_J$ is the
capacitance of a tunnel barrier at the NS interface. This
capacitance is neglected in our analysis and can be trivially
restored, if needed. Another -- more important -- difference is
that the frequency integral in Eq. (\ref{mpe}) cannot be extended
below the Thouless energy. This low frequency cut-off results in
the saturation of the interaction correction at small voltages and
temperatures. This effect is not contained in the standard
$P(E)$-theory which does not keep track of quantum coherence
of electrons entering an effective environment and which also does
not account for the sample geometry.

\begin{figure}
\includegraphics{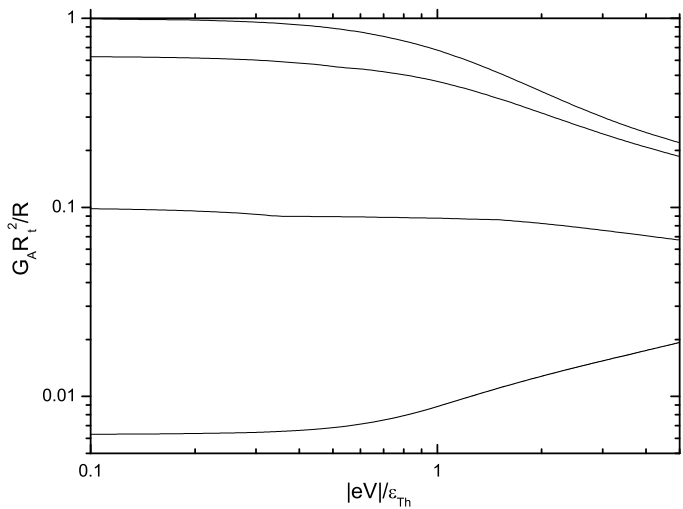}
\caption{The zero temperature differential Andreev conductance
$G_A=dI/dV$ of quasi-1d diffusive NS systems.  We have set
$RC\epsilon_{\rm Th}=0.01$. The dimensionless parameter $8/g$
equals to 0, 0.1, 0.5, 1.1 (from top to bottom). }
\end{figure}

In the limit $T \to 0$ the asymptotic expression for the  function $J(t)$  at
$t\gg RC$ reads
\begin{equation}
J(t)=\frac{8}{g}\left[\ln\left(\epsilon_{\rm Th} RC\right) -{\rm
ci}(|t|\epsilon_{\rm Th}) +i{\rm sign}\,t\,{\rm
si}(|t|\epsilon_{\rm Th})\right],
\end{equation}
where $ {\rm ci}\,t=-\int_t^\infty dz \cos z/z,\; {\rm
si}\,t=-\int_t^\infty dz \sin z/z$. In particular, for $t\ll
1/\epsilon_{\rm Th}$ we get
\begin{equation}
J(t)=-\frac{8 }{g}\left[\ln\left(\frac{e^\gamma t}{RC}
\right)+i\frac{\pi}{2} {\rm sign}\,t \right],
\end{equation}
where $\gamma \simeq 0.577...$ is the Euler constant. At longer
times $t\gg 1/\epsilon_{\rm Th}$ the function $J(t)$ approaches
the constant, $J(t)=-\kappa$, where
\begin{equation}
\kappa \simeq \kappa_1 =\frac{8}{g}\ln\frac{1}{RC\epsilon_{\rm
Th}}= \frac{8}{g}\ln\frac{e^2p_F^2 \Gamma \ln (L^2/\Gamma)}{\pi^2
v_F} . \label{11}
\end{equation}

The $I-V$ curve can now be easily evaluated. It is clear from Eq.
(\ref{mpe}) that in the limit of high voltages $eV \gg 1/RC$ the response
of an effective environment (normal metal) will be that of an
$RC$-transmission line. In this limit the interaction correction to the Andreev
conductance will behave analogously to that studied for single electron
tunneling within the $P(E)$-theory \cite{IN}. At lower voltages
$\epsilon_{\rm Th}\ll eV\ll 1/RC$ we find
\begin{eqnarray}
&& \frac{dI}{dV}=\frac{R}{2R_t^2}\sqrt{\frac{\epsilon_{\rm Th}}{|eV|}}
\Big[ e^{-\kappa}+
\\ && \frac{\sqrt{\pi}e^{-8\gamma/g}}{\Gamma(\frac{8}{g}+\frac{1}{2})}\left(
(2|eV|RC)^{8/g} -(\epsilon_{\rm Th} RC)^{8/g}
\right)\Big].\nonumber
\end{eqnarray}
In the limit of large conductances $g \gg 1$ this expression
can also be rewritten in a simpler form
\begin{equation}
\frac{dI}{dV}=\frac{R}{2R_t^2}\sqrt{\frac{\epsilon_{\rm
Th}}{|eV|}}e^{-\kappa}\bigg[1 +\frac{8}{g} \ln
\frac{|eV|}{\epsilon_{\rm Th}}\bigg].\label{f1}
\end{equation}
At smaller voltages $eV < \epsilon_{\rm Th}$ we again observe the
saturation of the differential conductance to the value
(\ref{satt}) where $\kappa$ is now determined by Eq. (\ref{11}).

The above results remain practically the same also in the presence
of screening electrodes in the vicinity of the sample. In that
case for $g \gg 1$ we again recover the logarithmic correction
(\ref{f1}) at large voltages while in the low voltage limit $eV\ll
\epsilon_{\rm Th}$ we find
\begin{equation}
\frac{dI}{dV}=\frac{Re^{-\kappa}}{R_t^2}\left[1+\frac{32}{15g}
\left(\frac{eV}{\epsilon_{\rm Th}}\right)^2\right].
\end{equation}
As before, in order to obtain the linear Andreev conductance $G_A(T)$ in
the regime $T \gg eV$ with the logarithmic accuracy, in the above
expressions it suffices to substitute $0.56 T$ instead of $eV$.

\begin{figure}
\includegraphics[width=8cm]{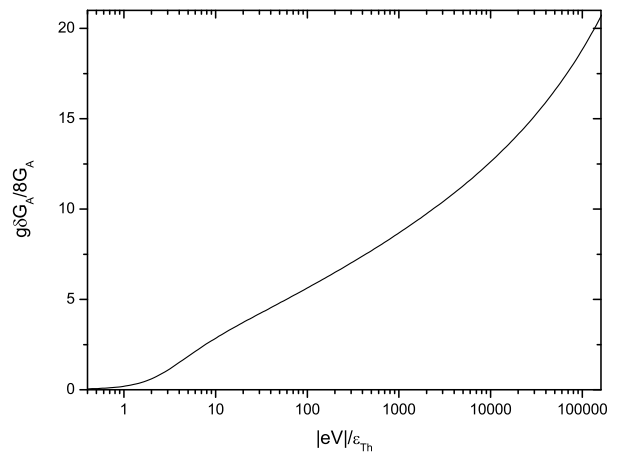}
\caption{The (normalized) value of the interaction correction to
the conductance $\delta G_A=G_A-G_A^{(0)}$ as a function of
voltage for $L_y=0.3L, L_z=0.1L$. One observes the saturation
regime at $eV \lesssim \epsilon_{\rm Th}$, the 1d logarithmic
regime (\ref{f1}) at $eV \gtrsim \epsilon_{\rm Th}$ which
eventually crosses over to the 3d regime (\ref{as3}) at $eV\sim
10^5 \epsilon_{\rm Th}$. }
\end{figure}

The differential conductance of quasi-1d diffusive NS structures
in the presence of electron-electron interactions is displayed in
Fig. 2 for several values of the parameter $8/g$. At larger
voltages one observes a competition between two types of ZBA --
conductance enhancement due to quantum interference of electrons
\cite{VZK,HN,Ben,Zai} and conductance suppression due to
electron-electron interactions. For small conductances $g < 16$
the second effect prevails and $dI/dV$ increases with voltage. At
large conductances $g > 16$ Coulomb effects become weaker and
$dI/dV$ shows the opposite trend. For smaller voltages $eV <
\epsilon_{\rm Th}$ the Andreev conductance saturates to the
voltage-independent value (\ref{satt}), (\ref{11}) for all values
of $g$. In this regime $dI/dV$ is always suppressed by
electron-electron interactions and this effect becomes stronger
with decreasing $g$.

Note, that the quasi-1d logarithmic behavior (\ref{f1}) is rather
robust and can manifest itself for a wide range of voltages even
if the sample length $L$ is not really much bigger than its width
and thickness. An example of such a behavior is presented in Fig.
3 for the sample with $L_y=0.3L, L_z=0.1L$. In this particular
case the logarithmic dependence of Eq. (\ref{f1}) remains
applicable within a very wide voltage range $\epsilon_{\rm
Th}\lesssim eV \lesssim 10^5\epsilon_{\rm Th}$, while the 3d
regime (\ref{as3}) may set in only at huge voltages $eV>10^5
\epsilon_{\rm Th}$.

\section{Current noise}

Our effective action analysis also allows to study fluctuations of
Andreev current in diffusive NS structures in the presence of
electron-electron interactions. Here we restrict our attention to
the current noise. The noise spectrum is expressed in terms of the
current operators $\hat I$ in a standard way as
\begin{equation}
{\cal S}(t,t')=\frac{1}{2}\left\langle \hat I(t)\hat I(t')+\hat I(t')\hat
I(t)\right \rangle-\left\langle\hat I \right\rangle^2.
\end{equation}
This expression can easily be translated into
\begin{equation}
{\cal S}(t,t')=-ie^2\left\langle \int_\Gamma d^2 r \int_\Gamma d^2
r' \frac{\delta }{\delta \varphi^-(\bm{r},t)} \frac{\delta
}{\delta \varphi^-(\bm{r'},t')}\delta S_A \right\rangle ,
\label{defnoise}
\end{equation}
where $\delta S_A$ is defined in Eq. (\ref{finalaction}).
Employing this general expression for the Andreev action together
with the definitions of current and noise (\ref{curdef}),
(\ref{defnoise}) one can derive the following relationship
\begin{equation}
{\cal S}(\omega,V,T)=e\sum_{\pm}\coth\left( \frac{\omega\pm 2
eV}{2T}\right)I\left(\frac{\omega}{2e}\pm V,T\right),
\label{nc}
\end{equation}
which remains valid irrespective of the dimensionality at
frequencies and voltages smaller than both the inverse charge
relaxation time $1/RC$ and the inverse elastic scattering time
$1/\tau$.

Note that the analogous expression between the current and the
noise spectrum was derived in Ref. \onlinecite{LL} for normal
tunnel barriers with interactions treated by means of the
$P(E)$-theory and in Ref. \onlinecite{SBL} within a more general
theoretical framework. Here Eq. (\ref{nc}) was derived for the
Andreev current and the Andreev noise spectrum in diffusive NS
structures in the presence of electron-electron interactions
treated within a microscopic theory. Eq. (\ref{nc}) has exactly
the same form as that \cite{LL,SBL}, one should only substitute
$2e$ instead of $e$ for the charge of the charge carriers. In the
zero frequency limit Eq. (\ref{nc}) reduces to the generalized
Schottky relation
\begin{equation}
{\cal S}(\omega,V,T)=2e\coth (eV/T)I(V,T),
\end{equation}
previously obtained in Ref. \onlinecite{PBH} in the absence of
interactions.

Combining Eq. (\ref{nc}) with the results derived in the previous
section one can fully describe the effect of Coulomb interaction
on Andreev current noise in diffusive NS structures. Hence, there
is no need to go into further details here.  Let us only mention
that -- similarly to the Andreev conductance -- at
voltages/frequencies below $\epsilon_{\rm Th}$ we expect shot
noise to be suppressed by the factor $\exp (-\kappa )$. At larger
voltages $eV > \epsilon_{\rm Th}$ and in the low frequency limit
the shot noise for quasi-1d NS structures should scale as
\begin{equation}
{\cal S} \propto |V|^{8/g-1/2}.
\label{pl}
\end{equation}

\section{Discussion}

The main observations of the present work can be summarized as
follows. In hybrid NS structures electron-electron interactions
generate two types of corrections to Andreev current. One of them
is just a voltage-independent renormalization of the $I-V$ curve
by the factor $\exp (-\kappa )$, while the other in general
depends on voltage and/or temperature. The net effect of
electron-electron interactions is always a reduction of the
Andreev conductance. This reduction approaches its maximum value
at temperatures and voltages below the Thouless energy of the
sample. In this low energy regime the tunnel barrier and a normal
metal act as a single coherent scatterer, whereas at energies above
$\epsilon_{\rm Th}$ they can be treated as independent ones in the spirit
of the $P(E)$-theory.

The magnitude of Coulomb effects essentially depends on the system
dimensionality and is characterized by the dimensionless parameter
$\kappa$. In 3d systems we find $\kappa \sim 1/p_F^2l^2 \ll 1$,
i.e. in this case Coulomb suppression of the Andreev conductance
is weak except for very disordered samples. In the 2d case
$\kappa$ is proportional to the parameter $1/p_F^2lL_z$ (which is
again small) but is enhanced by the two logarithmic factors (cf.
Eq. (\ref{22})), i.e. the net effect of electron-electron
interactions is not necessarily weak in this case.

This effect becomes even more pronounced in quasi-1d systems in
which case the value $\kappa$ is defined by Eq. (\ref{11}).
Obviously, the Coulomb suppression of the Andreev conductance can
be very strong for such systems. Combining Eqs. (\ref{ZBA}),
(\ref{ZBA0}) and (\ref{satt}) with Eq. (\ref{11}), and having in
mind that for typical values of the Fermi velocity in metals one
has $e^2/v_F \sim 1$, we arrive at a very simple estimate for the
Andreev conductance in the presence of electron-electron
interactions:
\begin{equation}
G_A/G_A^{(0)} \sim 1/N_{\rm Ch}^{8/g},
\label{estim}
\end{equation}
where $N_{\rm Ch} \sim p_F^2\Gamma$ is the number of conducting
channels in the normal wire. In Eq. (\ref{estim}) for simplicity
we have disregarded an additional weak (logarithmic) dependence on
$L$ and $\Gamma$, cf. Eq. (\ref{11}). Within this accuracy, Eq.
(\ref{estim}) demonstrates that Coulomb suppression of the Andreev
conductance in quasi-1d diffusive samples is determined by the
number of conducting channels to the power $8/g$. We believe this
prediction can be directly tested in future experiments.

Recently the authors \cite{Takayanagi} have experimentally
investigated the $I-V$ curves for hybrid systems composed of
multi-walled carbon nanotubes (MWNT) attached to a superconducting
electrode. The measurements \cite{Takayanagi} performed below the
superconducting critical temperature $T_C$ have revealed
power-law dependencies of the differential conductance on voltage
and temperature $dI/dV \propto T^{\alpha_S}$ and $dI/dV \propto
V^{\alpha_S}$. At smaller voltages $V < V_{\rm sat} \sim 1$ mV the
conductance was found to saturate to a constant value. Also at
$T>T_C$ the power-law dependencies have been
observed $dI/dV \propto T^{\alpha_N}$ with the value $\alpha_N$
somewhat smaller than $\alpha_S$.

It is interesting to compare our theoretical predictions with these
experimental results. The effect of conductance saturation
observed in Ref. \onlinecite{Takayanagi} just matches with our
prediction (\ref{estim}) and the value $eV_{\rm sat}$ appears to
be in a good agreement with the Thouless energy $\epsilon_{\rm
Th}$ estimated for MWNT \cite{Takayanagi}. Also the observed
power-law dependencies are consistent with our theoretical
picture. Identifying $\alpha_S=8/g-1/2$ and making use of the
experimental value $\alpha_S \simeq 1.28$, we can estimate the
effective dimensionless conductance for MWNT as $g \approx 4.5$.
Above $T_C$ for the superconducting electrode one also expects the
power-law dependence of the conductance on temperature
\cite{SZ,IN}. Neglecting the impedance of the bulk electrode one
would trivially get $\alpha_N = 2/g \approx 0.45$ which is
slightly smaller than the measured values $\alpha_N \approx 0.77$
close to $T_C$ and $\alpha_N \approx 0.55$ at higher temperatures.
It is clear, however, that above $T_C$ the impedance of the
superconducting electrode $R_S$ becomes non-zero and should also
be taken into account. Accordingly, the value $\alpha_N$ should be
modified as $\alpha_N = 2/g +2/g_S$, where $g_S=R_q/R_S$. This
additional contribution to $\alpha_N$ could further improve the
quantitative agreement between our theory and the experimental
observations \cite{Takayanagi}.

Finally, we would like to point out that it would also be
interesting to experimentally study the effect of Coulomb
interactions on the shot noise in quasi-1d hybrid NS systems,
like, e.g., ones investigated in Ref. \onlinecite{Takayanagi}.
According to the results obtained in Sec. VI, at low voltages one
expects suppression of the shot noise by the factor $\sim
N_{\rm Ch}^{8/g}$ while at higher voltages the power-law
dependence (\ref{pl}) should be recovered. Simultaneous current
and shot noise measurements could help to experimentally verify
the general relation (\ref{nc}) between the noise spectrum and
Andreev current in hybrid NS structures.

\centerline{\bf Acknowledgments}

We are grateful to D.S. Golubev for useful discussions. We also
thank the authors of Ref. \onlinecite{Takayanagi} for
communicating their results to us prior to publication.

\end{document}